# Yet another time about time …

## Part I: An Essay on the Phenomenology of Physical Time

Plamen L. Simeonov

> "I *now* predict that I was wrong."
> Stephen Hawking
> (McCarten, 2013)

**Abstract**

This paper presents yet another personal reflection on one the most important concepts in both science and the humanities: *time*. This elusive notion has been not only bothering philosophers since Plato and Aristotle. It goes throughout human history embracing all analytical and creative (anthropocentric) disciplines. Time has been a central theme in physical and life sciences, philosophy, psychology, music, art and many more. This theme is known with a vast body of knowledge across different theories and categories. What has been explored concerns its nature (rational, irrational, arational), appearances/qualia, degrees, dimensions and scales of conceptualization (internal, external, fractal, discrete, continuous, mechanical, quantum, local, global, etc.). Of particular interest have been parameters of time such as duration ranges, resolutions, modes (present, *now*, past, future), varieties of tenses (e.g. present perfect, present progressive, etc.) and some intuitive, but also fancy phenomenological characteristics such as "arrow", "stream", "texture", "width", "depth", "density", even "scent". Perhaps the most distinct characteristic of this fundamental concept is the *absolute time* constituting the *flow of consciousness* according to Husserl, the reflection of pure (human) nature without having the distinction between *exo* and *endo*. This essay is a personal reflection upon time in modern physics and phenomenological philosophy.

---

## 1. Prologue

Space and time are particularly modalities of human consciousness. Whereas space can be *real*ized through our eyes and limbs, time remains elusive to our minds and still bothers philosophers since antiquity (Dyke & Bardon, 2013). In the beginning of the 20th century science made a crucial switchover. Time was *spatialized*[1]. Until then the world was three-dimensional. With the Special Relativity Theory (Einstein, 1905; Minkowski, 1909) the concept of time became the fourth dimension[2] and an integral element of the new physical worldview. It caused confusion among many of Einstein's contemporaries, and not only with its relativistic dilation in the well-known equation with the speed of light $t' = t (1- (v/c)^2)^{½}$.

---

[1] It could be argued that time was spatialized earlier. Though time and space were separate for Newton (something that Einstein challenged), the clock time he assumed was laid out like space and was devoid of the dynamism of Bergsonian durée, (Bergson, 1912; Čapek, 1961).

[2] It was Laplace who treated time as just another dimension equivalent to space, and thereby paved the way for the development of relativity theory as it came to be understood (but not as Čapek, following Bergson, understood it).

This notion of relativistic time appeared wrong and unnatural, not without good reason, since:
- Time is *ineffable*, *eluding* science and mathematics: it can only be grasped through intuition and shown indirectly and partially through images (Bergson, 1912).
- Time is – concerning the extensities of the spatial – an *intensity*; it is a "qualitative, non-spatial, non-dimensional *inwardness*"[3] (Portmann, 1952).
- Time is a "heterogeneous dimension", which lies "transversely" to the "upright to it standing space dimensions".... "The 4$^{th}$ dimension is a 'dimensio sui generis'. Accurately phrased, it is not a 'di-mension', i.e. a dividing measure, but an *a-mension*, i.e. an element free from division and measurement.... a basic phenomenon without spatial character. It is a *quality*, whereas the measurability of the spatial dimensions lets them appear as quantities[4]." (Gebser, 1962).

In short, "time is not a categorical, and hence not a conceivable size[5]. It appears as an *acategorical* element, which cannot be registered systematically, i.e. spatially, but only *systatically*", when related to a combination or synthesis (systatics)[6] (Gebser, 2011, p. 51ff).

## 2. On time *with* time in physics: a personal perspective

Let us step back for a moment to realize a crucial moment in the history of physics preceding the development of both Special Relativity and Quantum Mechanics. This was the recognition of the limited frame of reference for physical laws, related to the fundamental issue of *time reversal* (T-symmetry) invariance and Loschmidt's Paradox, also known as the (ir)reversibility paradox. Newtonian mechanics, underlying low-level fundamental physical processes, is symmetric with respect to time reversal. "The equations of motion in abstract dynamics are perfectly reversible[7]; any solution of these equations remains valid when the time variable *t* is replaced by *-t*." (Thompson, 1874). However, the Second Law of Thermodynamics, describing the behavior of macroscopic systems, violates this principle. It determines a preferred direction of time's arrow by stating that the statistical state of entropy of the entire universe, taken as a closed isolated system, will always increase "over time" and never be negative. Loschmidt was provoked by Boltzmann's attempt to derive the Second Law from classical kinetic theory. He objected that it should not be possible to deduce an irreversible process from time-symmetric dynamics. Therefore, classical mechanics and statistical mechanics, both supported by a large body of evidence, were in conflict and hence the paradox. Today we know that certain physical forces (and laws) prevail at certain scales and become negligible at others; they may have evolved (Wheeler, 1983; Deutsch, 1986; Josephson, 2012) and cannot be derived from each other across the scales, at least to this moment[8]. We also know "now" that (physical) time is *asymmetric*, except for equilibrium states[9], where time symmetry holds. Generally, past, present and future cannot be exchanged in physics according to the Second Law.

---

[3] Die Zeit ist hinsichtlich der Extensitäten des Räumlichen eine Intensität, eine "qualitative, nichträumliche, nichtdimensionale Innerlichkeit". [translation from German and italics by the author]

[4] Die Zeit ist eine "heterogene Dimension", die "quer" zu den "senkrecht auf ihr stehenden Raumdimensionen" liegt... "Die 4. Dimension ist eine 'Dimensio sui generis'. Sie ist exakt ausgedrückt, keine "Di-mension", also eine einteilende Maßgröße, sondern eine A-mension, also ein von Maß und Messen befreites Element.... ein Grundphänomen, das kein Raumcharakter hat. Sie ist eine Qualität, während die Meßbarkeit der Raumdimensionen vornehmlich als Quantitäten erscheinen lässt. [translation from German and italics by the author]

[5] Time scales are *real*ized, i.e. experienced or felt as a *qualium* with no size at all (Salthe, 2013). Accordingly, time passing would be experienced at any scale 'the same', even though it is clear that when observing much larger or much smaller systems, they appear to be experiencing time more slowly or more rapidly than the observer's experience.

[6] "Sie – die Zeit – ist keine kategoriale und damit vorstellbare Größe. Sie erscheint als akategoriales Element, das nicht systematisch, also räumlich zu erfassen ist, sondern nur systatisch", d.h. in Bezug auf die Zusammenfügung der Teile zum Ganzen (Systase), "wahrnehmbar ist". [translation from German and italics by he author]

[7] However, motion itself is dissipative in the macroscopic case.

[8] The search for a General Unification Theory underlying all physical laws continues. Bohm's *hidden variables* (Bohm, 1952; Bohm, 1980) and Everett's *many-worlds interpretation* of quantum physics (Everett, 1957a/b; Everett, 1973) could still be instructive in this quest, not only for physicists – with a few exceptions they have been always claiming to address (replicable) phenomena in their research – but also for phenomenological philosophers.

[9] with possibly another "exception from the exception" for quantum noninvasive measurements which are supposed to violate time symmetry even in equilibrium (Bednorz et al., 2013).

The Special Relativity Theory brought essentially the *abolishment of simultaneity* [10]. Accordingly, it is impossible[11] to say in an *absolute* sense that two distinct events occur at the same time if those events are separated in space. Both Einstein and Poincaré came to the same conclusion explicitly referring to the concept of simultaneity (Einstein, 1905; Poincaré, 1898/1913). Before the arrival of Special Relativity space was considered a slice of reality cut by simultaneity, and (global) time required simultaneity to define valid relationships between objects in the universe. When simultaneity was abolished, both space and time were eliminated, at least in their classical meaning. If simultaneity is out, both space and time become meaningless as global parameters. Thus, physics rescued both concepts by assigning them local (relative) roles in a 4-dimensional space-time or time-space continuum. But in relativistic physics there are *no global space* and *no global*[12] *time* anymore. There is a *global space-time (or time-space)*, which is locally manifested as space and time. Abandoning only one of these concepts does not make sense[13] according to Special Relativity. Where there is space, there is time. They go hand in hand.

In fact the statements "time (or space) exists" and "time (or space) doesn't exist" imply something of a linguistic paradox. I am going to discuss this in some detail here, because it makes explicit an interesting cognition about the topic of this essay. This is the appraisal that different disciplines often use or borrow from each other's definitions and ontologies with different context to conceptualize their perspective upon a subject. We experience in our everyday lives that most misunderstandings between scientists and philosophers, but also among scientists and philosophers within their own domains, are due to this problem. A more systematic approach to cross- and intra-cultural *multilogue* (Bateson, 1972) is required.

I shall come now to the point. Existence (E) or non-existence (~E) is one of the most fundamental forms of human *experience*. Relativity theory postulates the absolute *existence* of space-time (or time-space). This projects in an observer's coordinates to existing space and existing time. But for an observer space exists "en-bloc" as thing/object (Gebser, 1962), while *time exists as a process*, *durée* (Bergson, 1912). This is an important distinction. A physicist adopting this wrong premise can state that relativity theory, which makes space and time observer dependent, eliminates existence as a fundamental concept. S/he may claim that existence requires a predefined concept of time, implying a temporal logic with operators such as 'before', 'after', 'always', 'never', 'eventually', 'sometimes', etc., in order to be applied within a postulate (suggestion or assumption of the existence, fact, or truth of something as a basis for reasoning). Mathematically it is not correct to describe spatial existence, temporal existence or the existence of a process (time) as one and the same thing[14]. In mathematics we have existence that is only logical and does not refer to either space or time. In the mathematical context we can discuss a process by adding a temporal operator or subscript, e.g. $\Box P$ or $P_\Box$ meaning "always P (holds)". But philosophically the situation is more complex. The issue is that *existence of processes* is a point of view of physicists[15]. But processes exist differently than non-temporal entities. One cannot just make logical arguments about relations that are mathematically ill defined.

---

[10] To claim this was the only way to reconcile with the constant speed of light in all inertial reference systems. Therefore it could be probably more correct to say that special relativity brought the *observer dependence*. [annotation by Felix T. Hong]

[11] In a way, the elimination of simultaneity by the Special Relativity is analogous to Heisenberg's uncertainty principle in Quantum Mechanics.

[12] The local concept of time was also discussed in earlier works of Salthe in connection with scale in the compositional hierarchy (Salthe, 1985; 1993), where he discussed small-scale moments nested within large scale ones involving the idea of 'cogent moment' for each scale. The local concept of time becomes also interesting when looking at it within a multi-agent system, i.e. in terms of mutually referenced 'change' relationships in concurrent multi-agent systems.

[13] mathematically and because most physicists tend to take space as given/existent

[14] [personal correspondence with Louis H. Kauffman, May 2015]

[15] Let's consider the existence of the Spacetime Manifold and call it M. In the mathematical models of physics M has mathematical existence. Physicists think of M as being "real". And they even think of how it may be embedded in higher dimensional spaces. This sort of reality is *beyond time* in the sense that we have time 'inside' M. So it would appear that *physicists believe in the reality of certain mathematical structures beyond spatial realities*. Of course they do. They believe for instance in the reality of the Hilbert spaces for quantum theory, quarks, strings, etc. We could imagine that these 'realities' can have physical correlates with given mathematical existences. [personal correspondence with Louis H. Kauffman, May 2015]

Mathematicians define mathematical structures to bring them into mathematical existence. But they cannot *define definition*. Nobody (except logicians!) can. That is, (most) mathematicians, but also scientists and philosophers, base their definitions on their ability to make *distinctions*. But *mathematicians cannot define what is a distinction!* Nobody can. What is a distinction is founded in human experience. *Mathematicians cannot define what is experience either!* It is the case that mathematical definitions come from human experience and are built in such a way that they can be exchanged like tokens of information. This is how we close the loop from science and mathematics back to phenomenology. In L.H. Kauffman's words: "Of course we can be as precise as possible about certain aspects of our experience. Thus I can tell you how to bake a cake, or give you the steps in a proof of a theorem, or tell you the complex steps that will lead you to observe the momentum exchanges that convince us of the existence of neutrinos. I can ask you to report your dreams. All sorts of relatively well-defined activities surround experiences that we have. But there will never be a logical mathematical definition of experience." Therefore, we should be happy to live in a formally incomplete world, which allows both space and time for experiencing anything. Nevertheless, for the purpose of discussing our example, let us adopt the convention that a process and an object are one and the same thing, despite the fact that the problem I will talk about is mathematically "ill-defined"[16]. It will only illustrate better the situation of a "compromised" communication. This special prerequisite is often neglected by philosophers and physicists and in no time we have a problem with time and other things. Let us have the physicist stating that experience (experience of existence) does not have any representation within the world described by Special Relativity. In other words, the relativistic view of the world eliminates the possibility of existence as a primary form of experience. Hence, the paradox of *(non-)existence of time*:

- o If time/space exists, $E(t/s) = 1$ or *True(E(t/s))*, then the proposition "time/space exists", is true: $t/s(E(t/s)) = 1$ or *True(t/s(E(t/s)))*, i.e. *True(E(t/s)) <=> E(t/s)*.
- o If time/space does not exist, $E(t/s) = 0$ or *False(E(t/s))*, the proposition "time/space does not exist", $t/s(E(t/s)) = 0$ or *False(t/s(E(t/s)))*, i.e. *False(E(t/s)) <=> ~E(t/s)*
    a) becomes *meaningless* for the physicist. This is indeed a very strong form of Aristotelian truth statement that physicists use to assess real-world situations[17]. If time and space do not exist in their *classical* meaning as absolute parameters, there is *no interdependence* between them. But they *exist* as local observer-dependent parameters according to the new meaning given to them by Special Relativity. Therefore, time and space *can* be interdependent in that sense. For a physicist's pragmatic logic, based on observable and provable facts if time or space does not exist, there is no notion of it, and hence we cannot build a proposition using it. So the claim is indefinite or meaningless[18].
    b) still *makes sense*[19] for a mathematician/logician, without respect to such parameters as space and time: if <whatever> does not exist, the assertion "<whatever> does not exist" is true in the sense of Tarski, (Tarski, 1995).

---

[16] Computer scientists use such pragmatic conventions. For instance, the programming language Scala addresses functions as software objects, to implement the extension of its underlying object-oriented language (Java) with elements of a general-purpose functional programming language (Haskell).

[17] We can assume that "reality" is a kind of "social construction", or more generally "construction". Salthe claims there is *no real world as such* [June, 2015, personal correspondence]. Indeed, in terms of his hierarchy theory (Salthe, 1985), this expanded "phenomenological" relation can be described as: {biological construction {social construction {individual construction}}}.

[18] What is assumed in this claim of physicists is that only a human being embedded in a reality with time can make sense of something. However, Kant, Spinoza, the neoplatonists and the Löbian machine (Marchal, 1990; 1992; 2013; 2015) believe that something can make sense independently of the presence or existence of a human being. From their viewpoint phenomenological philosophers and physicists are on the same anthropocentric side of reasoning.

[19] This proposition does not make sense in case that we imply some kind of a priori reasoning about the nature of space and time. Special Relativity adopts such kind of implicit knowledge as a convention, which can be expressed as a *second order logic*.

Consequently, the linguistic paradox[20] leading the physicist to the conclusion that there must be space (and time) cannot be seen as such by the mathematician, even if s/he adopts the convention of taking objects and processes as the same things. The reader is witnessing two different descriptions of the world, the one of the physicist and the one of the mathematician making a compromise. We should not forget that there is also another, third world view of the philosopher, which makes at least[21] three different types of phenomenology:

- phenomenology of the physicist → "pragmatic logic"
- phenomenology of the philosopher → "general (self-referential) logic"
- phenomenology of the mathematician → "machine (intelligence) logic"

All of these logics come from a person's different layers of *experience* (undefinable in mathematics), which cannot be perfectly matched. This is where all the troubles of interdisciplinary exchange come from. Most disciplines are closed in themselves and use local jargons of spoken languages (e.g. English). When their representatives speak to each other across and within disciplines, they usually understand different things. So, we cannot be precise enough when talking about something.

The reasoning of the physicist in the above example is basically the following[22]:

> *relativity = relative separation of space and time → existence requires concept of time (because of paradox) → existence becomes relative (a strange consequence) → because experience requires existence and/or existence is the most fundamental experience, relativity excludes these concepts as a matter of arbitrary[23] choice.*

The argument that if time or space does not exist, then the proposition that "time or space does not exist" becomes *meaningless* is not easy to comprehend by a non-physicist (because of its implicit second-order logic). This assumes that the meaning of a term is something that exists that it refers to. Even Bertrand Russell who argued for a referential theory of meaning would not go this far, but accept that one could refer to fictional objects such as unicorns or non-existent theoretical objects such as tachyons[24]. However, the relativistic worldview eliminates possibility of existence as a primary form of experience. Therefore, as long as experience and existence have the same equal ontological status as tachyons and unicorns in relativity, there is no problem with physics. But what does it mean to *exist*? Plato suggested that it could mean anything that can *causally influence or be causally influenced* by anything else, i.e. it can act something or be acted upon by something. This concept does not help much with categorizing time and space. According to Aristotle, there should be no time and space, because they are not causally connected to anything. However, it is also necessary to distinguish between what is in Aristotle's terminology *ousia*, or primary being, and what can only be understood as an *aspect of* a primary being. Thus, an atom can be considered as a primary being and existing as such (unless it is nothing but an intense curvature of space-time, in which case it only exists as an aspect of a primary being – spacetime). What might be meant is that *it* could be conceived to *exist by itself* irrespective of anything else. The question is what defines a primary being. Is this definition (in Aristotle's sense) supposed to be anthropocentric or not? If not, than the numbers 0 and 1 could be also taken for primary beings. This is e.g. the position of Digital Philosophy (Fredkin, 2003).

---

[20] In particular, 'exist' as an English (or basically, a Latin) verb, a form, which preferentially refers to an action, taking place *in time*. The existence of space-time fits poorly into this context. Thus, "the syntax of Korean has no true adjectives: they are a form of verb, so that a *red ball* becomes a *redness-expressing ball* and 'cogito ergo sum' becomes more like *thought happens, therefore I exist*. (To say that Koreans would start with na-nuhn, 'with reference to me'.) Therefore, discussions of existence are too often, unrecognized, discussions of syntax in one language or language family". [annotation by Tim Poston]

[21] We appear to be a priori more multiplied than with Everett's multiverse theory (Everett, 1957a/b; Everett, 1973). How to suppose an easy cultural understanding then? [personal correspondence with Bruno Marchal, May 2015]

[22] [personal correspondence with Marcin J. Schroeder, May 2015]

[23] Since all locales, events, and objects are unique, that means that all actual things are arbitrary.

[24] [personal correspondence with Arran Gare: June 2015]

Another illustration could be the status of space and time in Newtonian mechanics. For Newton, there would be space if there were no matter, and time if there were neither space nor matter. *Time* in this case would be the ultimate primary being, where space would be seen as existing and maintaining its existence through time. Matter is located in space and endures through time. Hence, when asking whether time exists, one could be arguing about whether Newton was right. However, one could also be asking about whether time is *experienced*, with a privileged 'now' with a past that used to be 'now', and a future that is very different from the past because it is not yet, but will become 'now'. This could be Husserl's very personal, experienced view on time. Time so conceived might not be a primary being, but rather an aspect of what there is, with reality conceived as *processes* that change in themselves and relate to each other. This kind of time is a *parameter accompanying the existence of the experiencer*. What is really being argued in the former case is: "Are there causal processes?". The latter is associated with a relational theory of time, which by rejecting Newton's theory of time, is sometimes taken to mean that time does not exist. However, if contrasted with Parmenides or some notion of the universe being a deterministic space-time block, as Einstein believed, and which many physicists believe, it could be taken as a defense of the *reality,* i.e. *existence,* of time. We will see later that reality itself can be regarded as a social (i.e. shared, at least 2nd person view) construct. In general, time always has also a surreal (i.e. individual, subjective) component.

If the second law of thermodynamics is considered to describe "real" (existing) phenomena, it is often taken to imply the "reality" (existence) of time and "time's arrow" as Prigogine argued (at this level[25] of description), although this is *time understood as an aspect of causal processes*. If the whole of "reality" can be grasped by some ultimate equations in which there is no reason to regard the relation between the present and the past and the present and the future – these are e.g. the cases of Newtonian mechanics and Maxwell's field theory – this can be taken to imply that time does not exist (at that level of description!), although what is being denied is time as Prigogine understood it.

The problem with the theories of relativity is the *absence* (or *uncertainty*) *of simultaneity,* and that what is taken to be a difference in time from a different reference frame, could be taken at least in part a difference in space. This is a very confusing detail, if time and space are considered as *primary beings*. However, if time and space are understood as *aspects of processes*, the issue is a lot less problematic, and can be interpreted to defend the reality (presence) of time. This is what Čapek argued (Čapek, 1961), claiming that in relativity theory there is never any questioning of the direction of causal influence, and the past is what can influence anything in the present, while the future is what can be causally influenced by what is in the present. The past and the future are then seen separated by a wedge of the 'elsewhere'. So, time is affirmed as a real aspect of what there is, while space as simultaneity is denied, and defined entirely in terms of time. Čapek argued that we should speak of time-space and not space-time, and Bohm followed him in this. However, it is generally accepted that Newton's framework of space and time is obsolete[26] today. The trouble remains that relativity theory and quantum theory are incompatible, as Einstein, Podolski and Rosen pointed out (Einstein et al., 1935) in order to bring quantum theory into question. Yet Aspect and his colleagues have shown with their "Bell test experiments" that, it is quantum theory that is validated, not relativity theory (Aspect, 1982a/b). They provided the proof that a quantum event at one location *can affect*[27] another event at a different location without respect to the distance and any obvious mechanism for communication between them.

---

[25] We will see this later in the discussion of the Wheeler-DeWitt equation.
[26] with yet such exceptions as the Hořava-Lifshitz gravity theory (Hořava, 2009) about short-distance interactions between non-relativistic gravitons, giving birth to the effective speed of light, Newton's constant and the cosmological constant altogether.
[27] This claim is not correct because: a) "can affect" is a causation statement, but correlation is not causation; and b) instantaneity is mistakenly read into these experiments as if the quantum entanglement was only between *simultaneous* states, but physicists have recently shown that *quantum correlation* is possible between photon pairs that never coexisted (Megidish et al., 2013).

With quantum entanglement, there is a sense that the entire universe is interconnected instantaneously. Yet, it is not clear of how this is to be interpreted, in the context of relativity theory, where the speed of light is the highest possible velocity in the universe. Gravity, in turn, cannot be quantized as the other known forces of interaction. In conclusion, it is not clear how space and time are to be interpreted as a consequence. Čapek's (1961) and Bohm's (1965) interpretations of relativity theory were remarkable in the past, but they may become obsolete, without a proper replacement. The perfect confusion continues today[28]…

… unless we allow the mediation of a *creative agency* (Deutsch, 2012; Josephson, 2015a/b). This puts the quest on a different ground that becomes apparent towards the end of this paper.

If experience is taken as a fundamental concept (e.g. in philosophical phenomenology), then relativity becomes a purely formal and practically useful description of the world, but useless for philosophical reflection. From the viewpoint of a pure abstract mathematician a physicist is also a kind of "phenomenological philosopher" putting the grounds of his/her reasoning not on contemplation and personal experience (the philosopher type of logic), but on pragmatic theoretical reasoning and (experienced) experimentation (using devices as extensions of human senses); both views are anthropocentric[29]. In contrast, the logic of the mathematician is universal[30]. Therefore if we speak of time, there must be some unspoken evidence (or convention) of its physical and psychical presence, without regard to how it can be defined. The trick with the substitution we made to represent a process as an object in order to carry out logical conclusions about the *(non-)existence of time*[31] at the same level of representation assumes a particular split[32] of space and time in the spacetime universe. It is generally not easy to find a common denominator for the individual phenomenological logics, which may also partly exclude each other. This field requires further exploration. The "problem of time" is well known in physics[33] and philosophy (Dyke & Bardon, 2013). The linguistic paradox I have discussed in detail here demonstrates only a small part of the reefs we can expect when steering a project in the inshore waters between such big islands of knowledge as physics, philosophy and mathematics. We will come back to it later when discussing a recent version of the Block Universe[34] Theory (Skow, 2005; Skow 2015).

Special Relativity challenged classical physics with its homogeneous and continuous structure of space and duration of time. The Newtonian concepts of space and time are linked, yet are independent of each other within the Aristotelian framework of object-in-space-(and-time)-before-subject (Rosen 2008a; Rosen 2008b). Space and time in everyday life constitute the classical arena in which physicists observe bodies in motion and measure (the pace of) changes they participate in via time-invariant mathematical equations. Precision, prediction and preclusion of hazards (causality) are good reasons in a pragmatic, extroverted world embossed by goals and virtues. Einstein's notion of *time's relativity*[35] was based on ideas of Hume and Mach (Norton, 2010). It banned the ether[36] from modern physics[37]. Yet philosophers like Bergson, Husserl, Heidegger and Merleau-Ponty continued to bother about time from a quite different perspective (Canales, 2015; Rosen, 2015).

Let us now open a new parenthesis and ask: what is time that we, human beings, experience? The mathematical equation for an ideal gas undergoing a reversible adiabatic process $pV/nT = R = const.$ is an example for the interdependence of parameters with different nature.

---

[28] [personal correspondence with Arran Gare, June 2015]
[29] They are equally social constructions. [personal correspondence with Stanley Salthe, June 2015]
[30] This should be considered as a general claim. Mathematicians can differ on logic, e.g. classical logic vs. intuitionist logic.
[31] As we will see later in the discussion of the Wheeler-DeWitt equation, this is a question of local/global or exo/endo relativism.
[32] A recent quantum field theory of gravity also suggested this comeback to a Newtonian nonrelativistic scheme (Hořava, 2009).
[33] http://www.wikiwand.com/en/Problem_of_time.
[34] also known as *Steady State Universe*. Relativity Theory and Newtonian Cosmology are usually taken as reference examples.
[35] *Cogent moment* (Salthe, 1993) is also a kind of relativity in time (dilaton); Pattee speaks of hierarchy *timescales* (Pattee, 1973)
[36] (Fresnel, 1819; Lorentz, 1899; Poincaré, 1904/6; Poincaré, 1906)
[37] However, Bohm suggested a more subtle idea of ether later (Bohm, 1952, 1980).

Planck's equation about quantum energy is another example: *const.= h = ET (E = hv = h/T)*. Whereas *T* denotes temperature in the former equation, it means time in the latter. Does temperature tell us something about the underlying molecular kinetics of the ideal gas? No. What tells us time in the second equation then? Deutsch uses the above example of a conservation law to demonstrate Wheeler's arguments that *symmetry and invariance are always indicative of a lower level of structure* than the one being described, while at the same time they mask the nature of this structure (Deutsch, 1986). This statement surprisingly reminds of Bohm's hidden variables interpretation of quantum theory (Bohm, 1952; Bohm, 1980) and Gödel's incompleteness theorems (Gödel, 1931).

Doesn't this appraisal of consistently lower levels of (related) physical reality down to molecules, hadrons and strings also apply for the equation of relativistic time dilation? Is not time also a kind of *intertwined* relation among entities of the underlying reality involving both the subject and the object, the container and the contained, the process and the processed? Isn't there a deeper level of description of the observable universe in which time doesn't appear at all[38]? Is it conceivable that time can only emerge in some sort of approximate description of reality by other means? Is it possible to estimate a *potentiality of time*? How many faces is time showing us? Are there different phenomenological interpretations of temporal relativity from the ones of the objectifying 3$^{rd}$ person descriptions of continuous space-time in Einstein's Special and General Relativity (Einstein, 1916) and fractal space-time in Nottale's Scale Relativity[39] (Nottale, 1993) inherent to the discontinuity thesis in quantum theory? Vrobel suggests to unify Merleau-Ponty's phenomenological notion of *nowness* as a 1$^{st}$ person's perception of *depth* (associated with individually registered superposition of events) involving the rationalization of systemic blind spots (Merleau-Ponty, 1945) and the one of temporal *length* ("cogent moment", Salthe, 1993) or duration (associated with the individually registered successive flow of events, Vrobel, 2013) from the perspective of the *obserpant*, the observer-participant who represents the model of the system itself (Vrobel, 2015). This approach is supposed to deliver an integral description of the subjective reality under investigation involving these two locally experienced (and possibly a few more) temporal scales ("timescales", Pattee, 1973) and *tenses* (Matsuno & Salthe, 2002) into one overall "fractal dimension" of superposed and sequenced events, known as *density* of time (Vrobel, 2007). These new scales of time are proposed to replace the single dimension in Einstein's relativistic model. However, even then, there must be still something elusively foundational remaining beyond such kind of indicative relativism in phenomenology. This evidence could be usually recognized in the method of integration or in the way of writing equations about the individual components of time. But integration techniques really eliminate the flow of time, since we know the beginning state and the final one; this is the Block Universe model! How would it be possible to allow for a comprehensible transition from one phenomenological domain of relevance to another (boundary conditions), as this is the case in Einstein's and Nottale's objective[40] relativities? Should we distinguish further fragmentations within and between phenomenological and non-phenomenological scales and descriptions, as this is the case with continuous and discrete functions in mathematics?

Phenomenological philosophy claims that time is not the place, the scene, the container or the medium for events (changes), nor a dimension along which everything flows (Desanti, 1968; 1975). According to Desanti, a French scholar of Husserl, we should forget "the ordinary meaning of the preposition "in" we spontaneously use when we talk about our experience of time. It is even this *use*, so ancient that should be the subject of our review. Really it would be strange that what we have learned to call "time" can contain anything. And yet we say without anxiety: "It is time that everything goes."

---

[38] Yet *not* with the cogent moment idea. [personal correspondence with Stanley Salthe, June 2015]
[39] Scale Relativity could possibly accommodate Salthe's "cogent moment" idea; this is actually not the case.
[40] "These are externalist concepts. A cogent moment is an externalist assessment of internalist experiences. Maybe that is why so many find it difficult." [personal correspondence with Stanley Salthe, May, 2015]

But what is happening "in" time *does not remain as a place.*[41] (Desanti 1992, p. 104). In fact, this is the major objection of Bergson against Einstein's Special Relativity (Canales, 2015), that he has dimensioned[42] time, something immeasurable in the same way as space, which is, of course (in everyday life), measurable[43]. This kind of reasoning in phenomenology is not that far from the one in modern physics.

> "There is a deeper problem, perhaps going back to the origin of physics... *time is frozen* as if it were another dimension of space. Motion is frozen, and a whole history of constant motion and change is presented to us as something static and unchanging… We have to find a way to *unfreeze time* — to represent time without turning it into space." (Smolin, 2006, pp. 256-257).

> "Today, the novelty that comes from quantum gravity is that space does not exist. … But combining this idea with relativity, one must conclude that the non-existence of space also implies the non-existence of time. Indeed, this is exactly what happens in quantum gravity: the variable *t* does not appear in the Wheeler-DeWitt equation, or elsewhere in the basic structure of the theory. … *Time does not exist*."[44] (Rovelli, 2012, p. 127, Kindle Ed.)

The claim about the *imaginary*, surreal, even exotic nature of time is not new in philosophy and physics (Mach, 1893; MacTaggard, 1908; Bairlein et al., 1962; Bedini, 1963; Barbour, 2001; Anderson, 2004; Rovelli & Vodotto, 2014). Of course, there have always been, too, physicists defending the *real* existence of time, even so real to define such a quantum variable as the *chronon* (Lévi,1927; Yang, 1947; Caldrola, 1980; Farias & Recami, 1997) with the idea in mind to reconcile special and general relativity with quantum field theory. This "atom" of time was supposed to be the duration for light to travel the distance of the classical (non-quantum) radius of an electron (Margenau, 1977). This model implies a lowest level of actuality, as asserted in the Planck scale.

In his book "Time Reborn" Smolin argues that physicists have inappropriately banned the reality of time because they confuse their timeless mathematical models with reality[45], (Smolin, 2013). His claim was that time is both *real* (which means *external* to him) and *fundamental,* hypothesizing that the very laws of physics are not fixed, but evolve *over time*[46]. This stance is not really a new one (cf. Wheeler, 1983; Page & Wootters, 1983). But it means again an absolute *external* reference axis and a direction for placing events in a sequence, which phenomenologists decline as the only option. Some of them, partly inspired by the late works of Heidegger and Merleau-Ponty, approach time neither from the standpoint of simultaneity alone, nor from that of succession. For instance, the dualism of these two concepts is surpassed in favor of a *temporal dialectic* in which simultaneity and succession are *entwined* (Rosen, 2008a/b), without denying their separate meanings. Heidegger's concept of "true time" speaks to this approach to phenomenology.

---

[41] "le sens ordinaire de la préposition 'dans' que nous utilisons spon-tanément lorsque nous parlons de notre expérience du temps. C'est même cet *usage*, tellement ancien, qui devrait faire l'ob-jet de notre examen. Vraiment il serait étrange que ce que nous avons appris à nommer 'temps' puisse contenir quoi que ce soit. Et cependant nous disons sans inquiétude: C'est dans le temps que tout se passe. Or ce qui se passe 'dans' le temps *n'y demeure pas comme en un lieu.*"  [translation from French and italics by the author]

[42] Salthe's "cogent moments" and Pattee's "timescales" are also dimensioning time. Yet, as being felt, they are immeasurable. Thus, we have again two perspectives on internalist time: the natural scientist's one and he the phenomenologist's one. Are they identical? Isn't Merleau-Ponty's *depth* and Vrobel's *density* time concepts describing the *methods* of realizing these hierarchies?

[43] Yet, even space is the result of gravitation according to General Relativity (Einstein, 1916).

[44] "Aujourd'hui, la nouveauté qui nous vient de la gravitation quantique est que l'espace n'existe pas. … Mais en combinant cette idée avec la relativité restreinte, on doit conclure que la non-existence de l'espace implique aussi la non-existence du temps. En effet, c'est exactement ce qui se passe en gravité quantique: la variable *t* ne figure pas dans l'équation de Wheeler-DeWitt, ni ailleurs, dans la structure de base de la théorie. … *Le temps n'existe pas*."  [translation from French and italics by the author]

[45] Did not Einstein do the same with the ether?

[46] a typical phrase, used by most physicists, often without being aware of its semantics.

In fact, Smolin appealed for a more philosophical approach to physics and to the issue of time (Smolin, 2006), but he *didn't* realize the necessity of a *phenomenological* approach. However, some scientists began to comprehend the importance of the phenomenological projection (Vrobel, 2015).

*Constructor Theory* seeks to express all fundamental scientific theories in terms of a dichotomy between *possible* and *impossible* physical transformations (Deutsch, 2012). Accordingly, a task *A* is said to be *possible* (*A*✓) if the laws of nature impose no restrictions on how (accurately) *A* could be performed, nor on how well the agents that are capable of approximately performing it could retain their ability to do so. Otherwise A is considered to be *impossible* (*A*✗). Deutsch argues that in both quantum theory and general relativity, "time is treated anomalously". He sees the problem in that time is not among the entities to which both theories attribute "objective existence (namely quantum observables and geometrical objects respectively), yet those entities *change with time*". Is not this an interesting appraisal? According to him "there is widespread agreement that there must be a way of treating time 'intrinsically' (i.e. as emerging from the relationships between physical objects such as clocks, e.g. Page & Wootters 1983; Barbour 2001; Barbour, 2012) rather than 'extrinsically' (as an unphysical parameter on which physical quantities somehow depend)." However, Deutsch reckons, it would be difficult to accommodate this in the prevailing conception, "every part of which (initial state; laws of motion; time-evolution) assumes that extrinsic status". According to Constructor Theory it is "both natural and unavoidable to treat both time and space *intrinsically:* they do not appear in the foundations of the theory, but are *emergent properties* of classes of tasks …". Note that the differentiation between "intrinsic" and "extrinsic" has different meaning than the one used in phenomenological philosophy.

Other projections, such as the *Circular Theory* (Yardley, 2010) and the *Structural Theory of Everything* (Josephson, 2015a) also question the laws underlying the foundations of physics. However, as long as they consider placing events "over time" extrinsically, they are all of the same kind: non-phenomenological[47]. (Note that *duration* has two definitions: endo and exo.) This is a very interesting conclusion in Steven M. Rosen's words: "The point may be that, in physics, time is treated 'extrinsically' because it does not lend itself to being formulated in objectivist terms, i.e., as something that is limited to the context of the objectified physical world."[48] Therefore, the terms "intrinsic" and "extrinsic" in Deutsch's objectivist usage are both "extrinsic" in the phenomenological sense: both are limited to a physicalistic paradigm that excludes the internal perspective of a *lived* (bodily) subjectivity or an agent-dependent reality (Rössler, 1987), a continuing durée (Bergson, 1912). This appears to be the key to all the "trouble with physics" (Smolin, 2006), and not only with physics[49].

---

[47] I dare to ask whether the laws underlying the foundations of physics are not purely phenomenological, and whether because of this subtle nature they have been escaping physicists to this moment. Signals for involving an introspective view on the creation of the universe can be recognized in (Barad, 2007), (Yardley, 2010) and (Josephson, 2015b). It is possibly worth venturing a contrasting juxtaposition of the articles in the following section of this special issue dedicated to mathematical and computational phenomenology (Hipólito, 2015; Kauffman, 2015; Ehresmann & Gomez-Ramirez, 2015; Marchal, 2015; Goranson et al., 2015), as well as their precursors in the 2013 special issue on Integral Biomathics (Goranson & Cardier, 2013; Marchal, 2013).

[48] [personal correspondence with Steven M. Rosen, May 2015]

[49] In their 2012 paper *No entailing laws, but enablement in the evolution of the biosphere* Longo, Montévil and Kauffman claim that biological evolution "marks the end of a physics world view of *law entailed dynamics*" (Longo et al., 2012). They argue that the evolutionary phase space or space of possibilities constituted of interactions between organisms, biological niches and ecosystems is "ever changing, intrinsically indeterminate and even (mathematically) *unprestatable*". Hence, the authors' claim that it is impossible to know "ahead of time the 'niches' which constitute the boundary conditions on selection" in order to formulate *laws of motion for evolution*. They call this effect "radical emergence", from life to life. In their study of biological evolution, Longo and colleagues carried close comparisons with physics. They investigated the mathematical constructions of phase spaces and the role of symmetries as invariant preserving transformations, and introduced the notion of "enablement" to restrict causal analyses to Batesonian differential cases (1972: "the difference that makes a difference"). The authors have shown that mutations or other "causal differences" at the core of evolution enable the establishment of *non-conservation principles*, in contrast to physical dynamics, which is largely based on conservation principles as symmetries. Their new notion of "extended criticality" also helps to understand the *distinctiveness* of the living state of matter when compared to the non-animal one. However, their approach to both physics and biology is also *non-phenomenological*. The possibility for *endo* states that can trigger the "(genetic) switches of mutation" has not been examined in their model. **All sciences are (still) externalist.**

Hawking's "Brief history of time" (Hawking, 1988) also conveys a purely objectivist-extrinsic model. His response to the firewall (or event horizon) paradox of the black hole (Almheiri et al., 2013) in early 2014 (Hawking, 2014) dug a hole in the black hole research (his own domain) opting for no event horizon at all. This was because quantum effects caused wild space-time fluctuations. In effect, the headline news in *Nature* and elsewhere was: *"There are no black holes."* (Meral, 2014). This announced the end of modern physics, as we know it. Time was placed somewhere "out there", away from the astronaut who was falling into a black hole. The theory speculated that the position of the event horizon is not locally determined. It is a *function of the future of the spacetime*" (whatever this may mean). In contrast, ontological phenomenology introduces a *dialectical blending* of "out there" and "in here", (Matsuno, 1995, 1998; Rosen, 2008a/b). Accordingly, time could be associated with the internal aspects of the *inner-outer* blend, while the astronaut falling towards the black hole (object-in-space) and experiencing the *now* constitutes the external aspect.

This is quite different from Einstein's twin paradox and means that we cannot simply relegate (special and general relativity) time to an external, merely objective, reality. Neither can we simply link it to the internal world of the obserpant. Rather, time is a manifestation of the *internal aspect of a reality* in which inner and outer are thoroughly *intertwined*[50] (superposed) (Rosen, 2015). Paradoxically enough, theoretical physicists have not discovered a simple trick of expressing the objectivist stance, or 3rd person description, in terms of (a) subjectivist one(s), or 1st person description(s), with the help of an ideal obserpant body size of *zero* (Vrobel, 2015), thus allowing for the 'exo' and 'endo' to become the same instance[51].

Sean Carroll, while being also an adversary of timeless physics, reasons that questioning the reality of time does not make sense, because *it* is "just there", i.e. it appears to exist, but maybe not as a fundamental or emergent property of the world (Carroll, 2010). In his opinion "the claim 'time does not exist' certainly doesn't mean the same kind of thing, as 'unicorns do not exist'"[52]. This argument seems reasonable: finally we speak of apples and oranges, especially if we are put in front of an audience of physicists, phenomenologists and mathematicians. This stance leads back again to the ideal gas law and Wheeler's argument about invariance and symmetry associated with lower levels of the (related) physical reality. The reader may recall the earlier discussion about Deutsch's reference to the conservation law of a reversible adiabatic process with an ideal gas ($pV/nT = R = const.$).

Why does time matter so much for us? There is something interesting as a corollary of all these viewpoints about time (even only in physics) put in the words of Sean Carroll[53]:

> "Temperature and pressure didn't stop being real once we understood them as emergent properties of an underlying atomic description. *The question of whether time is fundamental or emergent is, on the other hand, crucially important.* I have no idea what the answer is (and neither does anybody else). Modern theories of fundamental physics and cosmology include both possibilities among the respectable proposals.[54]"

---

[50] Rosen's work carries forward the phenomenological thinking of Heidegger and Merleau-Ponty, who sought to address the deep problem of subject-object dualism, a problem that haunts contemporary theoretical physics, particularly in the fields of quantum mechanics and quantum gravity (see Rosen 2008a/b).
[51] What is more appealing in this gedanken experiment is the possibility to investigate the *superposition* of multiple experienced (subjective) ideal body phenomena resulting in an observed/registered (objective) reality, an outcome that could be related to quantum computing and communication.
[52] Just because time does not appear in an equation does not necessarily mean that it does not exist. A novice can quickly refer to the equation for conservation of mechanical energy in the Bernoulli effect, which does not contain time because it is not needed for the derivation of this equation. But when he or she observes the action of a pumping hair spray, he or she also detects time in action: the liquid becomes atomized.
[53] http://www.preposterousuniverse.com/blog/2013/10/18/is-time-real/.
[54] The Big Bang and Big Crunch theories, for instance, imply that time must be real, since it emerges in the process of expansion of the universe.

Finally, Fotini Markopoulou argues that the problem of time is a paradox, stemming from an unstated faulty premise: the faulty assumption that *space is real,* (Markopoulou, 2008). She suggests that what does not fundamentally exist is not time, but space, geometry and gravity. Of course, this could be a valid assumption that nobody questioned, even phenomenologists. Why should not be space an illusion? In this case the quantum theory of gravity would be *spaceless* and not timeless. Markopoulou claims that if we are willing to throw out space from physics, we can keep time and the trade is worth it. But this goes back again to "Wheeler's loop" about symmetry and invariance cited above. This is because if time and space become mutually interchangeable, as Markopoulou implies, such symmetry suggests a more subtle principle deeper than either. Much of that knowledge should be comprehended metaphorically, "as a means to opening the mind to new insights" (Blake, 2003, p. 96). Thus, asking physicists "What is *real*[55]?" is a question that would please philosophers. Obviously, we can hypothesize about getting rid of either time or space.

> "There are two kinds of people in quantum gravity. Those who think that timelessness is the most beautiful and deepest insight in general relativity, if not modern science, and those who simply cannot comprehend what timelessness can mean and see evidence for time in everything in nature. What sets this split of opinions apart from any other disagreement in science is that almost no one ever changes their mind…" (Markopoulou, 2008).

Markopoulou's "time-for-space replacement" approach is extended in Shu's transformational "No Big Bang Theory" (Shu, 2010), a "Steady State" universe without beginning and end but alternating periods of expansion and contraction. It reminds of a partial solution of Alexander Friedman's dynamic ("oscillating" or "pulsating") universe, capable of expansion, contraction, collapse and emergence in a singularity (Friedman, 1922). Shu's idea is that time and space are not independent entities. They can be converted back and forth between each other. In his formulation of the geometry of spacetime, the speed of light is simply the conversion factor between the two. Similarly, mass and length are interchangeable in a relationship in which the conversion factor depends on both the gravitational constant G and the speed of light, neither of which need be constant. Thus, as the Universe expands, mass and time are converted to length and space and vice versa as it contracts. The behaviors of the different kinds of universes with peculiar interflections of the space and time depend presumably on their energetic (and informational?) status.

The confusion about the multiplicity of physical theories on the nature of time to this moment is almost complete. The situation appears to philosophers (incl. those reviewing this paper) even more frustrating. Almost 100 years after the Einstein-Bergson debate (Canales, 2015), the confrontation between the camps of physicists and phenomenologists does not look any different today. Why do we draw such parallels between the debates on time's validity as it occurs inside and outside of physics?

The first successful attempt to provide a unified picture of the physical world from micro and macro perspectives was made in the mid 1960s with the Wheeler-DeWitt equation (DeWitt, 1967). It combined the previously incompatible mathematical concepts of general relativity and quantum mechanics with the hindsight of a theory of quantum gravity. The price of this innovation was that time played no role in this equation of a "united universe" where nothing ever happens. The association with the equation for adiabatic extension of the ideal gas and Deutsch's clue on Wheeler's conclusion about the indicative nature of symmetry and invariance discussed earlier rise up again. This conundrum, known as "the problem of time" in physics, was resolved by Page and Wooters, who suggested a novel solution based on the idea of *quantum entanglement*[56] (Page & Wooters, 1983).

---

[55] To a nominalist reality is a social construction. [personal correspondence with Stanley Salthe, June 2015]
[56] See also footnote 27 on quantum correlation.

They suggested that a pair of entangled particles evolve in a way that reminds a kind of clock that can be used to measure changes. However, the result depends on *how* the observation is made. One way of doing this is to have an external observer outside the universe measuring the evolution of the particles using an external clock. In this case, Page and Wooters showed that this external observer would not detect any changes in the particles. Hence, time would not exist on a global scale (for God). But, if the observer[57] is located inside the universe, s/he would realize a change in the evolution of the particles, compared to everything else in the universe (for us, humans).

This is an important and powerful idea, suggesting that time is an *emergent phenomenon* because of the nature of entanglement. Until recently, it has been considered as a kind of mathematical-philosophical curiosity; physicists saw little chance of ever testing it.

In 2013 Moreva and collaborators delivered first tests of Page and Wootters' "emergent time entanglement" theory (Moreva et al., 2013). The experiment was about the creation of a toy universe consisting of only a pair of entangled photons with an observer able to measure their state in one of two modes: endo and exo. In the first case, one of the photons is treated like a clock with a tick that can alternate between horizontal and vertical polarization. An internal observer reading this clock (first photon) *will affect*[58] the polarization value of the second photon because of entanglement, thus becoming a part of the small universe, i.e. an *obserpant* in Vrobel's sense (Vrobel, 2013). The observer is then able to gauge the polarization value of the other (second) photon based on quantum probabilities. S/he then compares the measurement of the polarization of the first photon with the one of the second photon. The resulting difference is a measure of time. When passing through a thicker quartz plate photons experience different degrees of change. Repeating the experiment with plates of different thicknesses confirms that the second photon's polarization varies with time. Thus, the obserpant can measure the evolution of the system by becoming entangled with it. In the second case, s/she is outside the universe measuring the quantum state of the entangled photon pair and its evolution against an external clock, which is entirely independent of the toy universe. From that viewpoint, the state of the whole system is always the same, giving the appearance of a *static* universe where nothing happens; time does not emerge. Thus, the exotic hypothesis of Page-Wootters was placed on solid experimental grounds. Since physicists have recently become excited about the idea that gravity is an emergent phenomenon related to the thermodynamical aspects of gravity (Padmanabhan, 2009) and Verlinde's *entropic gravity*[59] theory (Verlinde, 2011), it is a relatively small step to presume that time could emerge in a similar way.

The experiment of Moreva's group is important, not only because it demonstrates how emergent gravity may take place, but also because it implies that quantum mechanics and general relativity are not so incompatible as initially assumed. The next step to be taken towards quantum gravity would be to scale up and make the transition to cosmology. It's one thing to show how time emerges for photons, and quite another to show how it emerges for larger objects such as galaxies. Viewed through the perspective of quantum entanglement and the works of Page, Wooters and Moreva's team, there is indeed no "problem of time" with Wheeler-DeWitt's "time-excluding equation". However, not everyone agrees that this is the correct route to unification of the quantum and classical worlds. For instance, Smolin and Mangabeira argue that every correct description of the Universe *must* include time, even at the expense of dethroning mathematics as the oracle of sciences (Smolin, 2013; Mangabeira & Smolin, 2014).

---

[57] observer-participant, *obserpant*, in Vrobel's sense (Vrobel, 2013)
[58] Again, correlation is not causation (cf. footnote 27). A measured polarization value of the first photon can have implications about measurements of the second (in the past, present or future), but *no effect* on them.
[59] http://www.wikiwand.com/en/Entropic_gravity.

Another recent theory, also grounded on quantum entanglement, explains energy dissipation in terms of quantum information and communication rather than statistical law distribution used in thermodynamics so far. It could also possibly give us the answer to the fundamental question about the source of the irreversible arrow of time, and hence about the nature of time itself. In their 2009 article, Linden and colleagues demonstrate that energy disperses and objects equilibrate, because of the way elementary particles become *intertwined* during their interactions (Linden et al., 2009). Thus, physical systems reach equilibrium, or a state of uniform energy distribution, within an infinite "amount of time"[60] by becoming *quantum mechanically entangled with their surroundings*. Does not this sound similar to the stance of ontological phenomenology? This connection becomes perhaps clearer when taking into account the role of the observer in QM. Obviously, elementary particles cannot be conceived in purely objective terms: the entanglement of particles is inseparable from the entanglement of observer and observed. In phenomenological terms, this corresponds to the entanglement of subject and object underlying Merleau-Ponty's temporal *depth* dimension, which can be related to Heidegger's notion of *true time*.

Hence, the unidirectional time's arrow can be said to "emerge" from Heidegger's "true time" notion, with the latter put in relation to QM[61]. This suggests an identification of a completely entangled sub-Planckian world with the profoundly intertwined world of Merleau-Ponty's "brute Being" (1969, p. 211).

These results certainly resonate with earlier work on quantum information (Lloyd, 2007). It becomes clear that as objects interact with their surroundings — as, for instance, the fluid particles in a hot cup of coffee collide with the air — information about their properties leaks out and becomes dispersed over the entire environment. This local information loss causes the state of the coffee to stop changing ("in time"?), even as the pure state of the entire room continues to evolve, except for rare, random fluctuations. Consequently, a cold cup of coffee does not spontaneously warm up. Hence the continuous *increase of entropy* stated in the second law of thermodynamics is interpreted to be the result of the *loss of information through quantum entanglement*. Thus the appraisal puzzling generations of physicists that the arrow of time does not seem to follow from the underlying laws of physics, which work in the same manner when going forwards and backwards "in time", has probably found its physical explanation "without time[62]". Today, Linden, Lloyd and their colleagues regard the arrow of time (process) differently and reckon that information becomes increasingly dispersed, but it never disappears entirely. They believe that *even if entropy increases locally, the overall entropy of the universe remains constant at zero*. This is a very interesting claim, in particular with respect to the emergence and evolution of life, mind and consciousness as related to quantum entanglement with the surrounding environment.

The standard explanation for "time's arrow" experienced in our expanding universe has been that it is an emergent property of thermodynamics' Second Law imposed at special initial conditions. Barbour and colleagues demonstrate that this is not the case. They suggest a cosmological hypothesis about the existence of a "parallel universe with backwards time" that has emerged at the Big Bang simultaneously with ours (Barbour et al., 2014). The idea goes back again to the finding that nearly all known laws of physics look exactly the same without regard to time's direction. The researchers used the Newtonian n-body problem as the simplest nontrivial time-symmetric law to model a dynamically closed universe. They studied a computer simulation of 1000 point-like particles interacting under gravitational force.

---

[60] Again, do we need this quantization of time or can we not get rid of it? Don't we speak of a process?
[61] Similarly to energy, we could regard time as an extended complex valued variable, such as a quaternion (Chappell, 2012). This assumption could lead us to a unified theory of quantum theory and thermodynamics, which would let thermodynamics "emerge" already at the microscopic level instead of only at the macroscopic one.
[62] But one could still ask: "Doesn't loss of information also take time to occur?"

The system's dynamic behavior was investigated using a measure of its complexity, corresponding to the ratio of the distance between the system's closest pair of particles and the distance between the most widely separated particle pair. Because of the special properties of the system all typical solutions fall into two groups at a uniquely defined point. In each one of them, the measure of complexity varies, but grows irreversibly between rising boundaries from that point. During this process indicated with complexity growth, structures storing dynamical information are created and act as "records" of the universe history. The authors have shown that essentially every configuration of particles would evolve into a low-complexity state analogous to the Big Bang. Each solution can be regarded then as having a single past and two distinct futures emerging from it. For an internal observer (obserpant) situated in one half of the solution and aware of the records only of that part of the expanding universe, there will be a unique past and a unique future direction ("time's arrow") deducible from the available records. The obtained results from the computer simulation are considered by Barbour and colleagues to be a proof of (another relativistic) principle that all the solutions of a time-symmetric dynamical (global) law have strongly time-asymmetric (local) behaviors for internal observers. This interesting outcome could be also regarded as a special instantiation of Everett's multiverse theory (Everett, 1957a/b).

Philosophy has been a permanent companion of science over the ages. In the past century, thinkers like Kuhn and Popper have contributed a lot for the development of scientific thought and standards (e.g. Popper, 1935; Kuhn, 1962). In particular, Karl Popper was favoring radiation over entropy increase as the source of irreversibility and the arrow of time (Popper, 1956; 1957; 1965; 1967) simply because he regarded Boltzmann's statistical theory unacceptable. His predilections were not proven to be correct: the real source of time's asymmetry is currently considered to be cosmology and gravitation (e.g. Barbour et al., 2014). However, the mere fact of scrutinizing scientific questions from a number of different perspectives established as "the" *touchstone* for the scientific merit of a theory its *falsifiability*, or *refutability*, or *testability*. It was accepted and adopted by generations of scientists up to the present day. This method is challenged now. Physics has been increasingly seduced in favor of more demanding theories which could hardly offer falsifiability.

An innocent bystander could get the impression that all these efforts to explain anything in the world, incl. the nature of time and consciousness, by means of QM are in effect attempts to aggrandize the entire universe into one big quantum formulation. Ancient philosophers and future generations could rightly ask us today: "Is this ambition to make the world generally as equally ununderstandable as QM a way to make QM more palatable?"

Should not we worry about, deliberate on and debate this increasing monocultural orientation tendency in physics, and not only in physics, in the past few decades? Should not we recognize the possibility that some quantum field phenomena beyond electromagnetism (e.g. strings) could have been thought out and *constructed* (engineered) by way of experiments matching adopted and developed formalisms to explain them? They are *not discovered* in the way e.g. the new distant galaxy EGS-zs8-1or the double helix structure of the DNA. Is there any evidence for their existence in the *act*ual world, outside the smart mathematics of the bold theories and the machinery of the experiments with anticipated results? Should not we look for alternative theories and explanations of the Universe? Perhaps we should afford asking these provoking questions now, before the crisis of physics (Smolin, 2006; Frank & Gleiser, 2015) deepens too much. On the other hand, we know well there cannot be a paradigm change without a crisis (Kuhn, 1962). Thus, crisis is good.

"QM engineering" has been an expensive venture. If it does not justify the investments soon, with the detection of supersymmetry as experimental proof of the beautiful M/string theory, physics may have to be redefined as scientific discipline. But we need to keep the same standards for all sciences without exception, don't we?

It is clear that in order to complete the transition to a Unified Theory and understand how it works the link to general relativity remains to be made. Why should we put all the eggs in just one basket? "Men seem inclined to react to a problem either by putting forward some theory and clinging to it as long as they can (if it is erroneous, they may even perish with it, rather than give it up), or by fighting against such a theory, once they have seen its weaknesses." (Popper, 1963). Therefore, these dialectic processes of confrontation and development are normal. Popper also remarked that the dogmatic approach of sticking to a theory as long as possible is of significant importance. "Without it we could never find out what it is in a theory – we should give the theory up, before we had a real opportunity of finding out its strength."

The question about the nature of time still remains one of the greatest riddles of physics. We cannot say if quantum theory helped any further to lift the curtain on it. But something needs to be done about it. Maybe physicists should begin studying *time perception*[63] and Husserl's phenomenological concept of internal *time-consciousness* (Husserl, 1991; Heidegger, 1996; Varela, 1999; Kortooms, 2002).

> "But we must not forget that all things in the world are connected with one another and depend on one another, and that we ourselves and all our thoughts are also a part of nature. It is utterly beyond our power to measure the changes of things by time. Quite the contrary, time is an abstraction, at which we arrive by means of the change of things; made because we are not restricted to any one definite measure, all being interconnected." (Mach, 1893).

\* \* \*

The goal of this paper was to reaffirm that time duration is an abstraction (Mach) and to appraise internal time-consciousness (Husserl). The remaining hard issue is how to integrate these two aspects into a single coherent body for research. If one starts with Husserl, what should be answered first is the question of how experiences at large could already be *latent* in time-experiences of the individual. Once this has been clarified, the next step should be to identify the nature of an *agency* for the succeeding abstraction along Mach's line. Maybe the notion of *delays* could bridge these two projections. For instance, we could think of them as blind spots in the obserpants that have been fully compensated regarding their simultaneity (Vrobel, 2015). I will come back to this later after discussing another approach.

The different aspects of abstract unbound time and internal conscious time can be explored with Skow's adaptation of the Block Universe theory[64] (Skow, 2005, 2015) mentioned earlier. Accordingly, time does not pass or move inside space. It is not frozen at one omnipresent moment either. Skow reckons that time should be regarded as a dimension of spacetime being part of the basic fabric of the universe, which is what relativity theory holds. However, objects in this model are considered not located at a single moment, but spread out "in time", just in the way they are spread out "in space"; hence the term "Block Universe". Skow believes that events do not happen once to vanish forever. They simply *exist* on par in different parts of the spacetime, but there is no way to link the same "timed" objects, which means no time travel.

This model of the Block Universe was extended with the Moving Spotlight Theory (MST), which allows that the past and the future *exist* on a level with the present. Accordingly, spacetime has a temporal metric and a temporal orientation with instants of time pictured as points without extensions or as the "tips of backward-facing light cones, possibly softened in some quantum or twistor way: local in all four dimensions"[65].

---

[63] http://www.wikiwand.com/en/Time_perception; The Experience and Perception of Time: Stanford Encyclopedia of Philosophy, http://plato.stanford.edu/entries/time-experience/.
[64] to be dated back to Minkovski's 4D spacetime model (1908) and McTaggart's B-Theory of time (1908).
[65] [personal correspondence with Tim Poston, July 2015]

However, only one moment (at a time) is regarded as absolutely present, and that moment keeps changing, as if a spotlight were moving over it.

The Moving Spotlight Theory is said to be compatible with Special Relativity (Skow, 2008). There is also a mirroring between the temporal structure of spacetime and the structure of *supertime*, a dimension similar to time, but not time; it is said to be distinct and disconnected from it and makes the transition to the relativistic theory. Skow did not mention phenomenology, but his attempt to render "scattered" time existence within relativism appears still sufficiently self-consistent, even allowing for an extension towards the "moving spotlight" version, to be interesting for research on the boundary between physical and psychical phenomena.

One could speculate further about how to integrate the different endo and exo relativistic models of time, also with respect to a possible realization of a perceptual demonstrator as an AI device or an ALife organism (Smith, 2012). But what can be said about a unified approach to exophysical and endophysical quantum concepts of time, such as e.g. the (cor-)relation between (shared) non-local experiences and geomagnetic fluctuations, a physical phenomenon that phenomenological philosophers have not considered (to my knowledge) in their research yet?

Let us come back to the Vrobel type of fully compensated delays to bridge abstract and experienced time mentioned six paragraphs above. It is interesting to note that there is ongoing research on *physical time delays* by Koroatev and colleagues (Korotaev et al., 2003; 2005) confirming the results of earlier work by Kozyrev's team on the experimental investigation of the properties of time (Kozyrev, 1971; 1980). They have found delayed correlations between solar activity and geoelectromagnetic patterns with long time delays. Their results basically boil down to macroscopic nonlocality, a correlation of different dissipative processes without any local carriers of interaction (Korotaev et al. 1999; 2000; 2002), which can be regarded as an example of a "physical expanded *now*" on a large scale.

These objectified time delay phenomena were also registered as consequences of geomagnetic storms[66] entering hidden portals[67] in the geomagnetic field causing magnetic reconnection[68] and the so-called "Sedona effect" correlations (Miller & Lonetree, 2013) between geomagnetic anomalies, EEG brainwaves and Schumann resonances[69]. This suggests a dependency that philosophical phenomenologists have not considered in their research so far: that human beings and other inhabitants of the planet Earth are most probably subjected to such *cosmic delays* too and that our sensations, moods, thoughts, theories, abstractions are also influenced by them[70]. Geomagnetic disturbances can disrupt cell phone services, damage satellites and knock out power grids. But also our physical complaints do not leave any doubt about such a relationship (Greenwald, 2015). Biophysicists know well that our bodies are not Faraday cages and human consciousness cannot be regarded as an isolated phenomenon. Yet, they may also err in their logic, in particular with respect to biosemiotics (Kull, 2015).

Let us come back again to the interpretation of time in relativity theory, which I would like to comment from a bit different perspective. Minkowski's idea of a space-time continuum (1909) merged the three real spatial dimensions with an *imaginary* one of time $ict$, with a basic unit imaginary producing the correct spacetime distance $x^2 - c^2t^2$, thus providing an explanation for Einstein's special relativity theory in terms of the structure of space and time.

---

[66] disturbances of the Earth's magnetosphere caused by a solar wind shock wave and/or a solar magnetic field cloud interacting with the Earth's magnetic field.
[67] https://www.nasa.gov/mission_pages/sunearth/news/mag-portals.html.
[68] a fundamental physics process in which magnetic field lines come together and explosively realign, causing particles in the high-density magnetic field area to be accelerated close to the speed of light.
[69] http://www.wikiwand.com/en/Schumann_resonances.
[70] [personal correspondence with Susie, Vrobel, 14. May 2015]

In 2012 Chappell and colleagues propose an alternative to a planar Minkowski space-time of two space dimensions and one time dimension (Chappell et al., 2012), by replacing the imaginary complex number unit $i$ with the Clifford bivector $i=e^1 e^2$, the square of which is also minus one, but can be included without the addition of an extra dimension, as it is an integral part of the real Cartesian plane with the orthonormal basis $e^1$ and $e^2$. The authors reckon that with this model of planar spacetime, using a 2D Clifford multivector, the spacetime metric and the Lorentz transformations follow immediately as properties of the underlying algebra. This also leads to momentum and energy being represented as components of a multivector. Chappell and colleagues provide an efficient derivation of Compton's scattering formula, and a simple formulation of Dirac's and Maxwell's equations. Based on the mathematical structure of the multivector, they produce a semi-classical model of massive particles, which can then be viewed as the origin of the Minkowski spacetime structure and thus a deeper explanation for relativistic effects. The authors also claim to have found "a new perspective on the nature of time, which is given a precise mathematical definition" as the bivector of the plane. They should be kindly reminded that a richer and more elegant mathematical model of time was proposed over 40 years ago by Charles A. Musès based on his hypernumber concept, which he elaborated throughout his lifetime (cf. e.g. Musès, 1972; 1977; 1983; 1985; 1994). This work was cheerfully ignored, although there is practically no excuse for missing the discussion of such essential references by experts in the field (incl. the reviewers at PLOS ONE) today. Unfortunately, such cases of superficial recherché (if not worse) practices are quite frequent with an increasing tendency in modern scientific literature, despite easier access to digital content and search engine optimization.

Thus, authors publish sometimes redundant or less significant works finding lame pretexts in the frantic principle "publish or perish". Why does this happen? Why did producing and accelerating publications (at any price!) became the dominating criteria for success in science? Is it possible to avoid "happy ignorance" and "innocent borrowings" of older ideas without paying tribute to the original authors? Is not time to introduce more stringent quality criteria beyond speedy peer reviewing, impact indexing and ranking digests, which are obviously insufficient methods for assessing scientific excellence in the modern "Big Data" driven world? Is not time for a paradigm shift *from quantity to quality* in science? Would not be better if authors are given *more time* to think, study, discuss and work on a theme?

* * *

I hope that my contemplations "left the vast darkness of the subject unobscured" to the reader, as a friend reported what Bateson told him about Whitehead's response to a talk on quantum mechanics. He added that the "gift" character of time's *now* cannot be appreciated enough, neither in science, nor in philosophy: "The *now* is epiphany. Time is dream-substance. And *personalized* at that." Heidegger noted that we can only say of Being *es gibt*, "it gives"[71]. And — by way of emphasizing the intimacy of Being and time — he suggested that time is the horizon of Being. Beyond Merleau-Ponty's concept of (temporal) *depth* (Merleau-Ponty, 1945) and Vrobel's concept of (temporal) *density* (Vrobel, 2015), phenomenological time certainly has also intrinsic *color*, *sound*, *scent*, *taste* and *texture* in a moment of joy or sorrow, as personal as it can be. Is it so important to us, because we all (including philosophers) are mortal? How do we know what we know?

---

[71] Heidegger took the unusual step of working with the non-idiomatic, literal meaning of *es gibt* ("there is"): "it gives." He did this to emphasize that Being or "true time" entails an active process of *giving*. This is where the gift ("das Geschenk", „die Gabe") comes from.

## 3. Epilogue

What I realized when studying the diverse threads of thought concerning time on both sides of the divide between physics (as representative of science) and phenomenological philosophy – but also within physics itself – is that the major differences between the diverse viewpoints, compacted in what each side is missing to understand the other, are those of Batesonian kind: the little bits of anticipated information that each discipline and each opinion imply as known and shared (axioms). "What we mean by *information* — the elementary unit of information — is *a difference which makes a difference*, and it is able to make a difference because the neural pathways along which it travels and is continually transformed are themselves provided with energy. The pathways are ready to be triggered. We may even say that the question is already implicit in them." (Bateson, 1972, p. 336). In other words, we always need to know some second order logic, and presuppose a second order of "order" (cybernetics) usually shared within a distinct community, to realize what a certain claim, hypothesis or theory means. In Matsuno's opinion[72] Bateson's phrase "must be a prototypical example of second-order logic in that the difference appearing both in the subject and predicate can accept *quantification*. Most statements framed in second-order logic are not decidable. In order to make them decidable or meaningful, some *qualifier*[73] needs to be used. A popular example of such a qualifier is a subjective observer. However, the point is that the subjective observer is not limited to Alice or Bob in the QBist[74] parlance". This is what I wished to emphasize when trying to understand the different viewpoints in logic of mathematicians, physicists and philosophers in the dispute about the existence of time in the beginning of this essay. An essential aspect of this "implicate order" (Bohm, 1980) can be seen in the grammatical formulation of theses such as the law of motion:

> "While it is legitimate in its own light, the physical law of motion alone framed in eternal time referable in the present tense, whether in classical or quantum mechanics, is not competent enough to address how the now could be experienced. … Measurement differs from the physical law of motion as much as the *now in experience* differs from the present tense in description. The watershed separating between measurement and the law of motion is in the distinction between the now and the present tense. Measurement is thus subjective and agential in making a punctuation at the moment of now." (Matsuno, 2015).

In particular, the distinction between experiencing and capturing experience of time in terms of language is made explicit in Heidegger's *Being and Time*: "… by passing away constantly, time remains as time. To remain means: not to disappear, thus, to presence. Thus time is determined by a kind of Being. How, then, is Being supposed to be determined by time?" (Heidegger, 1969/1972, p. 3). Matsuno's comment on this is: "Time passing away is an abstraction from accepting the distinction of the grammatical tenses, while time remaining as time refers to the temporality of the durable now prior to the abstraction of the tenses." (Matsuno, 2015). Therefore, when trying to understand the "local logics/phenomenologies" of the individual disciplines (mathematics physics, philosophy, etc., incl. their fields), one should be aware of the fact that the capabilities of our scientific language (usually English) are not limitless: "the *now* of the present moment is movable and dynamic in updating the present perfect tense in the present progressive tense. That is to say, the now is prior and all of the grammatical tenses including the ubiquitous present tense are the abstract derivatives from the *durable now*", (Matsuno, 2015). This presupposes the adequacy of mathematical abstractions specifically invented or adopted and elaborated for the expression of more sophisticated modalities of time's *now* than those currently used in such formalisms as temporal logic (e.g. TLA, Lamport, 1994; 2002). There is much to be explored in what other languages and cultures deliver to physics (e.g. Peat, 1996).

---

[72] Koichiro Matsuno on FIS (http://listas.unizar.es/cgi-bin/mailman/listinfo/fis, 27.06.2015).
[73] italics by the author in this paragraph.
[74] Quantum Bayesianism: http://www.wikiwand.com/en/Quantum_Bayesianism.

Finally, let us discuss shortly the epistemology of time in physics. Stepney (2014) investigated interesting aspects of local vs. global physical and computational time related to "Newtonian" or differential ($t + dt$) and "Lagrangian" or integral ($\int L\, dt$) least action models of motion in classical physics. The discussion is carried along Wharton's argument (Wharton, 2012) claiming that (classical von Neuman - Turing) computation used today is essentially Newtonian iterative calculation[75], whereas the (quantum) universe (e.g. Stapp, 2007) at its heart is most naturally described in a least action Lagrangian formulation, and hence the universe is not a computer in the classical sense. Wharton's argument is of course not correct in Feynman's context (Feynman, 1982; 1985) regarding computing for physics, which ultimately led to a whole field of research on quantum computation. Certainly, any other kind of unconventional computing relevant to Stepney's discussion (neural networks, reaction-diffusion chemical systems, cellular, DNA, field and optical computing, etc.) is also not adequate to Wharton's interpretation. The question, however, which modes of computation (and hence, implicitly, which aspects of *time* used in them) are appropriate in naturalistic physical models and simulations is still valid. This certainly involves the definition of initial and boundary conditions. However, they appear nondeterministic outside the Newtonian scheme. The universe is a complex evolving system characterized by always emerging and disappearing elements and properties, which are traditionally handled from a third person perspective in physics. Phenomenology does not play a practical role in physics today. How can we define an adequate model of the physical universe then? Is it possible to ultimately comprehend it or we can only capture some fractal, "local" aspects of what the universe *means* to us as individuals participating a community with shared knowledge?

Hankey's article in this special issue is addressing the interesting aspect of how to find a complexity basis for phenomenology through information states at criticality points (Hankey, 2015). It becomes apparent that the issue of time requires integrated phenomenological treatment in physics. Most physicists, however, such as David Ritz Finkelstein and Gordana Dodig-Crnkovic[76] do not see any reason to introduce phenomenology into physics. They consider *endophysics* (Rössler, 1987; 1998) as "just fine" and "objective" (inter-subjective) third-person account of the introspective aspect of physical systems. In particular Dodig-Crnkovic regards cognition as the way for capturing *agency* in the course of the subject-object interactions. She considers phenomenology as a second order phenomenon capturing the experience of the individual observer rather than the object under observation. Dodig-Crnkovic reckons that physics is about agency, not about experience, which belongs to the domains of psychology, philosophy and art. I will come back again to Hankey's model.

Stepney was, however, pursuing a different question while wishing to understand computation as physical phenomenon: "Can we *make* our computations more physics-like?" Her answer was "Yes" when addressing unconventional computational architectures based on physical processes. However, asking the reverse question of whether mathematics and computation in the broad sense (and not physics!) can be regarded as the foundations of our physical universe appears also valid. Thus, Marchal claims that arithmetic, or *any* first order logical specification of any Turing universal system, is the foundation of everything, mind and matter, (Marchal, 2004; 2013). He demonstrates that this logical conclusion follows from the hypothesis of computationalism and a minimal amount of Occam razor reasoning. Refuting Marchal's argument appears to be a real challenge for both physicists and mathematicians, if they take it seriously. But we discussed earlier that logical incompatibility between the different disciplines makes such efforts difficult (see also Kull, 2015).

---

[75] The problem with assuming that all physics can be calculated using iterative methods applied to differential equations (second order PDEs) is that Norbert Wiener's Cybernetics showed that systems with feedback require integro-differential equations for their elucidation, which introduce immense complications when trying to solve them. The nonlinearities in critical phenomena are examples in this respect. Therefore, we possibly need new different mathematical approaches for capturing such phenomena.
[76] [personal correspondence]

Nevertheless, physicists like Stepney and Dodig-Crnkovic argue that physics, which is the world that *acts*, is primary and not its representation in an agent such as a human's mind in the form of arithmetic, symmetry or any second-order phenomenon that emerges[77] from that substrate. This is again a fundamental epistemological hallmark. At the end it is not the question of who is right or wrong in this discussion about the foundations of our physical universe, but where a scientist or philosopher feels at home and what he or she considers as primary. If a mathematician reasons at the abstract level of symmetries and logic then he or she can express everything else in the universe in terms of logic and arithmetic. A physicist, however, could be interested to understand how exactly arithmetic and computation emerge from a living cognitive substrate, but he or she would not question the existence of the physical world *outside* of that substrate! Philosophically, both positions belong to the two opposite traditions of the Platonic and Aristotelian interpretations of the world. There is no final resolution in the favor of one or the other. It is a dynamic, dialectic unity of opposites, akin to the discourse about free will and its relations to upward and downward causation discussed by Felix Hong in the complement to this special issue (Hong, 2015) and its predecessor article on the role of pattern recognition in scientific research (Hong, 2013). We know more if we accept both viewpoints for what they are and look for their synergies.

Yet, there is something new in distributed computation from a physicist's perspective. Newton and the modern era physicists did not think in terms of networks to this moment. Whereas classical physics discusses the one-, two- and many-body problems, complex networks with multiple feedbacks (Simeonov, 2002) introduce a new (emerging) phenomenon, in particular with nonlinear criticalities, and interconnected time structures could be different. This is also an essential characteristic of living systems, which I am going to discuss in a sequel article.

Time is obviously a phenomenon at the interface between (human) mind and the physical world. We may not be able to answer completely the fundamental question we set out to, but we can acquire important insights in many fields of crucial importance for humanity if disciplines go hand-in-hand when dealing with this concept and continue the dialogue. The first step beyond the frame of mathematics and natural sciences was made in this special issue extending our hand to phenomenological philosophy, which has been considering an entirely different approach to time and other dualistic (endo and exo) phenomena in the past 500 years.

Our belief is that both fields, science and philosophy, can mutually enhance each other again when working on a common language/logic, as it was the case a few centuries ago. We should regain our lost ability to understand each other because of the increased fragmentation of knowledge.

While being recognized as physicist, Galileo Galilei was also probably among the first phenomenologists of time in history. As the story goes, when being a student of medicine at the University of Pisa, he visited the cathedral at Vallombrosa and observed a lamp hanging from the ceiling swaying in perfect rhythm. The young scholar checked immediately his own *pulse* to ascertain his observations and was fascinated that the lamp took the same amount of *time* to swing, no matter how large the range of swinging. Being a son of a music teacher, he must have had a quite good sense of rhythmic movement. Curiously enough, Galileo used these insights later in his studies on pendulums and *clocks*.

We need to devise a way of integrating knowledge, reasoning and intuition like Rennaissance and "Vitruvian" men of all ages did (Capra, 2008; 2013).

---

[77] But how do we know that the physics itself as we know it is *not* such a second-order emergent phenomenon?

It was not my intention to present a satisfactory response to the question *"What is time?"* in this essay. Rather, I wished to emphasize that the basic challenge we have at hand is that of integrating a number of different descriptions *from* and *between* different disciplines, as well as *within* the disciplines themselves, as shown here in the case with physics. What we need is to understand these concepts and their domains, to know their adequate frame of reference in order to correctly apply them to other domains. In particular, we should become able to capture and represent the subtle nature of time in the ideas of Gebser (1962) and Portmann (1952) we began this essay with. At the same time we should try to overcome the paradoxical limits of science reflected in a vivid illustration about the substantiality of the quantum domain given by Brian Josephson:

> "Nature [in this regime] behaves rather like an insect that responds to our attempts to swat it, utilizing a determination of just where it will be at the appropriate future instant, by changing its position. This ensures that our prediction will be incorrect, with the result that the insect survives our attempts to dispose of it."
>
> (Josephson, 2014)

Hankey's work on the realization of self-organized criticality in this issue (Hankey, 2015) could provide some basic clues on the way. Backed by previous research in a number of disciplines (e.g. Back et al., 1987; Holland, 1992; Penrose, 1994; Kauffman, 1996; Chalmers, 1997; Back, 1999; Hameroff & Penrose, 2014) the author hypothesizes about this phenomenon as an "almost inevitable" *mode of operation* of complex biological systems, allowing the maintenance of switching processes close to criticality to avoid unwanted switching blockages. Hankey reckons that this improves the sensitivity of the switching processes, i.e. the *response-ability,* of living systems to environmental changes, which are in the core of adaptation, fitness and survival. This mode of operation is supposed to represent *a form of optimized regulation* corresponding to *system 'health'*. Hankey's model of circulating information round the criticality loop of a living system supports the realization of a *sense of self* incorporating the intrinsic properties of phenomenological time: i) the immanent sense of continuity in time passing, of Heidegger's 'Being in Time', and ii) the inherence of time in experience as interpolation between, and connection of (possibly emotionally stressed and modulated) successive events.

I hope that the Integral Biomathics approach (Simeonov, 2010; Simeonov et al., 2012; Simeonov et al., 2013) will help us advancing fast in the desired direction.

This work is by far from being finished even in its "Part I". A number of interesting models and theories addressing the "problem of time", such the diverse approaches to quantum gravity (Unimodular Gravity, Bigravity, Dilaton Gravity, Massive Gravity, F(R) Gravity, Hořava-Lifshitz Gravity, etc.), were omitted to simplify the presentation. The interested reader may wish to consider the ones related to *quantum local times* (Kitada, 1994; Kitada & Fletcher, 1996; Jeknić-Dugić et al., 2014), *classical and quantum time crystals*[78] (Shapere & Wilczek, 2012; Wilzcek, 2012) and *timeless quantum vacuum* (Caligiuri and Sorli, 2013; Sorli, 2014; Fiscaletti & Sorli, 2014a/b). Other knowledgeable research was not addressed at all in this paper for the lack of *timespace* available to me while co-editing the contributions to this special journal issue. However, I wish to honor this inspiring work in my digest about the phenomenology of physical time in case that the interested reader wishes to explore more about these concepts: Paul Davies (Davies, 1995), Huw Price (Price, 1997; 2002), George Jaroszkiewicz (Jaroszkiewicz, 2000; 2001) and Henry Stapp (Stapp, 2007). The key essence of these ideas could help us reviving *Natural Philosophy* as a discipline and lead us towards the 'grand answer' that the reader may have expected to see at the end of this paper.

---

[78] analogous to ordinary crystals in space, representing spontaneous emergence of a clock within a time-invariant dynamical system.

In the hope that the reader had a good *time* reading this essay, which I finish *now*[79]

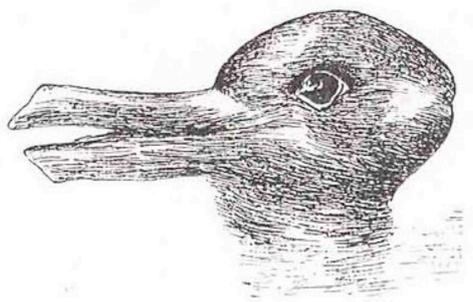

with a last citation and the aspiration for a sequel in a couple of years:

> "The principles … that I have proposed may be false. … But if the idea as a whole is false, *something else*[80] will have to remedy the deficiencies of the prevailing conception." (Deutsch, 2012)

<div style="text-align:right">PLS, Berlin, in July 2015</div>

**Acknowledgements**

I wish to thank to my mother Rossitza for her love, dedication, patience, support and inspiration for writing this paper and all throughout my life. I am deeply grateful to Steven M. Rosen, Louis H. Kauffman, Arran Gare, Koichiro Matsuno, Marcin J. Schroeder, Susie Vrobel, Otto E. Rössler, Ursula Saar, Andrée C. Ehresmann, Bruno Marchal, Tim Poston, Brian D. Josephson, Gordana Dodig-Crnkovic, Felix T. Hong, Stuart Kauffman, Denis Noble, Shaun Gallagher, Ted Goranson, Howard Pattee and Stanley S. Salhe for their stimulating feedbacks and recommendations, which qualify them as co-creators of this incredibly dense and wired essay. Wikipedia articles were of great help when trying to verify references and historical dependencies bewean the numerous concepts and theories used in this paper.

---

[79] borrowing an illustration from Wikipedia (source: http://www.wikiwand.com/en/Paradigm_shift) to indicate a cryptic conjecture left to the reader.

[80] According to Deborah Mayo (1996) Thomas Kuhn should be also interpreted that way: "For Kuhn, in a genuine science, anomalies give rise to research puzzles. In our recasting of Kuhn this becomes, in a genuinely scientific enquiry, anomalies afford opportunities for learning, opportunities for learning from error. The aim of science is not avoiding anomaly and error, in our opinion. The aim is being able to learn from anomaly and error." So, why should not we accommodate such ideas as post-mathematical science (Josephson, 2015b) and phenomenological philosophy (Rosen, 2008b) in Integral Biomathics (Simeonov, 2010; Simeonov et al., 2012; 2013)?